\documentclass[sigconf,authorversion,nonacm]{acmart}

\usepackage{multirow}

\AtBeginDocument{%
  \providecommand\BibTeX{{%
    \normalfont B\kern-0.5em{\scshape i\kern-0.25em b}\kern-0.8em\TeX}}}

\copyrightyear{2020} 
\acmYear{2020} 
\setcopyright{acmlicensed}
\acmConference[ICMI '20]{Proceedings of the 2020 International Conference on Multimodal Interaction}{October 25--29, 2020}{Virtual event, Netherlands}
\acmBooktitle{Proceedings of the 2020 International Conference on Multimodal Interaction (ICMI '20), October 25--29, 2020, Virtual event, Netherlands}
\acmPrice{15.00}
\acmDOI{10.1145/3382507.3418817}
\acmISBN{978-1-4503-7581-8/20/10}

\newcommand{\sugg}[1]{\textcolor[rgb]{0,0,0}{#1}}
\begin{document}


\title{Studying Person-Specific Pointing and Gaze Behavior for Multimodal Referencing of Outside Objects from a Moving Vehicle}


\author{Amr Gomaa}
\affiliation{%
  \institution{German Research Center for Artificial Intelligence (DFKI)}
  \city{Saarbr{\"u}cken}
  \country{Germany}
}
\email{amr.gomaa@dfki.de}

\author{Guillermo Reyes}
\affiliation{%
  \institution{German Research Center for Artificial Intelligence (DFKI)}
  \city{Saarbr{\"u}cken}
  \country{Germany}
}
\email{guillermo.reyes@dfki.de}

\author{Alexandra Alles}
\affiliation{%
  \institution{German Research Center for Artificial Intelligence (DFKI)}
  \city{Saarbr{\"u}cken}
  \country{Germany}
}
\email{alexandra_katrin.alles@dfki.de}

\author{Lydia Rupp}
\affiliation{%
  \institution{German Research Center for Artificial Intelligence (DFKI)}
  \city{Saarbr{\"u}cken}
  \country{Germany}
}
\email{lydia_helene.rupp@dfki.de}

\author{Michael Feld}
\affiliation{%
  \institution{German Research Center for Artificial Intelligence (DFKI)}
  \city{Saarbr{\"u}cken}
  \country{Germany}
}
\email{michael.feld@dfki.de}


\begin{abstract}
Hand pointing and eye gaze have been extensively investigated in automotive applications for object selection and referencing. Despite significant advances, existing outside-the-vehicle referencing methods consider these modalities separately.
Moreover, existing multimodal referencing methods focus on a static situation, whereas the situation in a moving vehicle is highly dynamic and subject to safety-critical constraints. In this paper, we investigate the specific characteristics of each modality and the interaction between them when used in the task of referencing outside objects (e.g. buildings) from the vehicle.

We furthermore explore person-specific differences in this interaction by analyzing individuals' performance for pointing and gaze patterns, along with their effect on the driving task. Our statistical analysis shows significant differences in individual behaviour based on object's location (i.e. driver's right side vs. left side), object's surroundings, driving mode (i.e. autonomous vs. normal driving) as well as pointing and gaze duration, laying the foundation for a user-adaptive approach.
\end{abstract}

\ccsdesc[500]{Human-centered computing~User studies}
\ccsdesc[500]{Human-centered computing~Pointing}
\ccsdesc[500]{Human-centered computing~Gestural input}
\ccsdesc[300]{Human-centered computing~HCI theory, concepts and models}

\keywords{Multimodal interaction; pointing gestures; eye gaze; head pose; object referencing; personalized models}


\maketitle

\begin{figure}[!t]
	\begin{center}
		\includegraphics[width=0.9\linewidth]{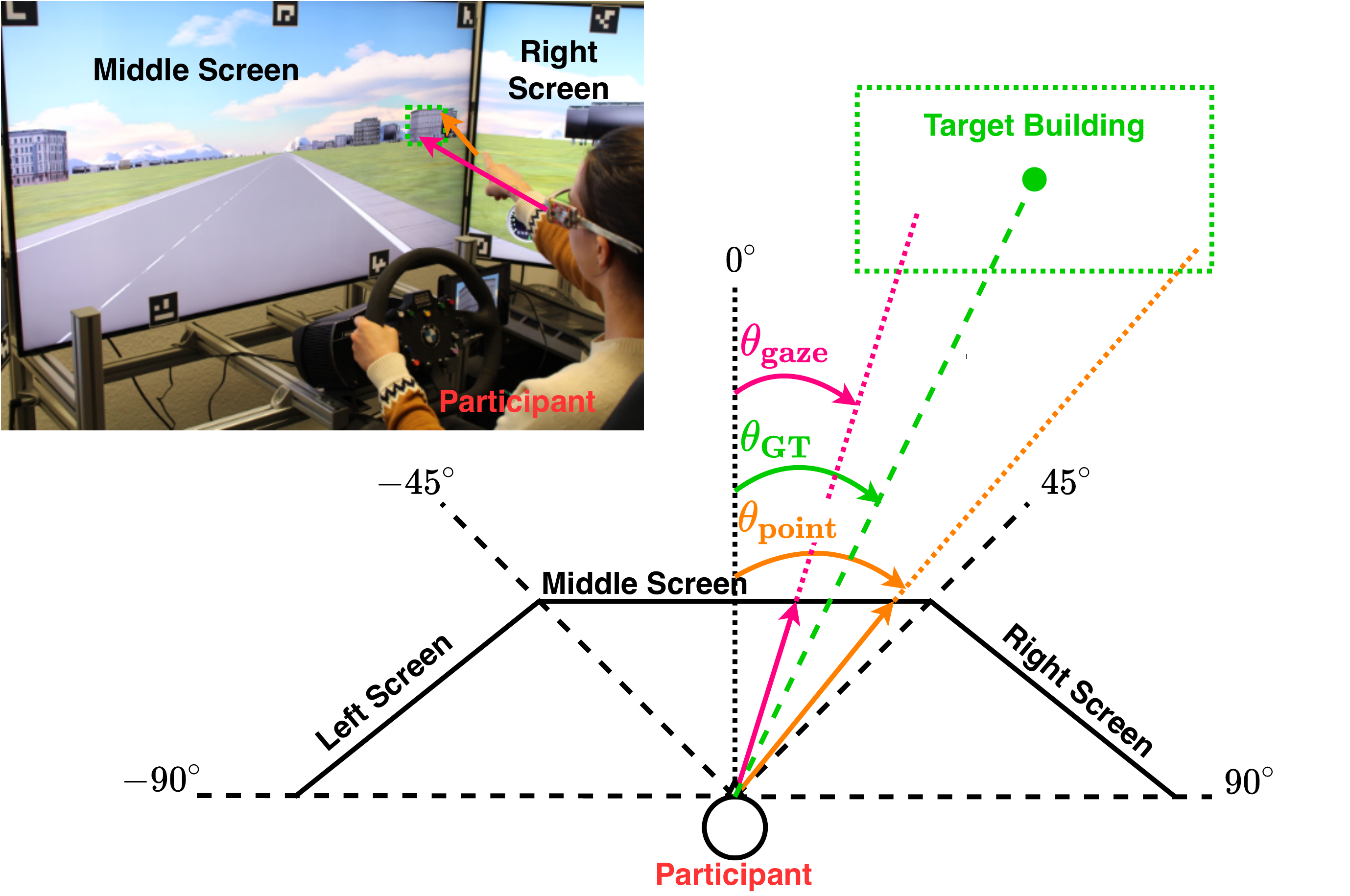}
	\end{center}
	\caption{Referencing objects outside a moving vehicle. Pointing and gaze vectors are transformed into a common 1D cylindrical coordinate system and compared to the ground truth vector.}
	\label{fig:implementationoverview}
\end{figure}

\section{Introduction}

As cars get more intelligent and their functionalities develop exponentially, the need for a natural form of interaction increases. There are several advantages to using hand gestures, eye gaze, head movements and speech over traditional touch-based interaction methods, such as increased simplicity and naturalness when interacting with a relatively complicated machine like a modern car, in addition to a reduction in distraction during the primary task (i.e. driving)~\cite{pickering2007research,Garber2013,feld2016combine,Roider2017}. Thus, researchers have tried to incorporate these modalities to control various components inside the vehicle~\cite{molchanov2015hand, molchanov2015multi, ohn2014hand,zobl2003real,roider2018see,sezgin2009multimodal}. 

Furthermore, the advances in gesture recognition and computer vision domains allow for real-time interaction with the car's surrounding environment, especially for referencing landmarks and buildings outside the vehicle using hand pointing and eye gaze gestures~\cite{rumelin2013free,fujimura2013driver,kang2015you}.  
Despite these major advances and their application in the automotive domain, existing methods often rely on single modality solutions~\cite{vora2017generalizing,vasli2016driver,ohn2014hand} or only partially use the second modality (i.e. as a control channel only)~\cite{nickel20043d,poitschke2011gaze}, which is insufficient for accurate referencing.
Although the use of multimodal fusion could outperform single modality approaches, fusion still suffers from multiple challenges~\cite{baltruvsaitis2018multimodal,atrey2010multimodal} such as representation (e.g. exploiting complementarity and redundancy), alignment (e.g. time synchronization), translation (e.g. transforming coordinate systems) and co-learning (e.g. knowledge transfer) which are addressed within our approach. There have been several attempts for modality fusion between eye gaze and symbolic hand gestures. 
However, to the best of our knowledge, neither multimodal referencing of objects outside a moving vehicle using eye gaze and deictic pointing gestures nor the interaction between them has been studied before.
Besides, since pointing and gaze behaviour when referencing objects differ greatly among users~\cite{brown2016exploring,nickel20043d}, a global solution which fits all users is not feasible~\cite{brown2014performance} and a person-specific approach would perform better.

This paper makes the following contributions: In a medium fidelity driving simulation, we investigated the multimodal behaviour for the secondary task of referencing objects outside a moving vehicle. Through a quantitative and qualitative analysis, we studied pointing and gazing differences in behaviour when performing this secondary task under several conditions (e.g. when the object was near or far from the driver). Furthermore, we analysed individual differences among users to increase understanding of personal pointing and gazing behaviour. Finally, our approach better adheres to the dynamic and safety aspects of the driving process, as it is tested with a long driving scenario.

\section{Related Work}

Pointing and eye gaze gestures have been studied rigorously in multiple applications. Nickel et al.~\cite{nickel20043d} laid the foundation for pointing gestures in a human-robot interaction environment, Jing et al.~\cite{jing2013human} used pointing gestures in selecting objects on large displays and Kehl et al.~\cite{kehl2004real} studied pointing using both arms. Similarly, Vidal et al.~\cite{vidal2012detection,vidal2013pursuits} studied eye gaze interaction with moving targets on a large display, while Cheng et al.~\cite{cheng2018smooth} studied across devices eye gaze interaction. However, these methods studied pointing and eye gaze in a stationary, standing environment that is not applicable in a driving scenario.

Furthermore, Roider et al.~\cite{Roider2017} and Nesselrath et al.~\cite{feld2016combine} studied the selection of objects inside the vehicle using hand gestures, eye gaze or speech commands separately. Similarly, Poitschke et al.~\cite{poitschke2011gaze} studied referencing objects inside the vehicle using eye gaze gestures while Sezgin et al.~\cite{sezgin2009multimodal} studied selection using speech commands and facial recognition. Recently, Roider et al.~\cite{roider2018see} also assessed the combination of pointing and passive eye gaze to reference objects inside the vehicle. However, the use cases of these approaches were simple two-to four-object classification ones that are hard to extend to a generic outside-the-vehicle referencing approach like the one presented in this work.

Moreover, referencing objects outside the vehicle has been investigated using different approaches and modalities. Rümelin et al.~\cite{rumelin2013free} used free-hand pointing gestures, Fujimura et al.~\cite{fujimura2013driver} used hand-constrained pointing gestures, Kang et al.~\cite{kang2015you} used eye gaze gestures, while Kim et al.~\cite{kim2014identification} and Misu et al.\cite{misu2014situated} used speech-triggered head pose trajectories. However, these studies focused on single-modality approaches that were lacking in performance. For example, eye gaze suffered from the Midas touch problem and sporadic involuntary eye movements that hindered accurate tracking~\cite{young1975survey,rayner2009eye,land2009looking,Moniri_2018} while pointing gestures suffered from performance inconsistency among users~\cite{brown2016exploring,brown2014performance,nickel20043d}. 

Although these studies are not directly comparable with our approach due to previously mentioned differences, they still presented insights into differences in pointing and gazing behaviour among users that were used in our analysis. For example, Rümelin et al.~\cite{rumelin2013free} and Nickel et al.~\cite{nickel20043d} reported an average pointing time of 1.8 seconds. Furthermore, Rümelin et al. defined three phases for driver's gazing (i.e. glancing) behaviour during this pointing time as follows:
\begin{itemize}
    \item \textbf{Information Glance:} Users look at the object to select
    \item \textbf{Pointing Position:} Users point at the object and quickly draw their eyes back to the road while keeping their hands pointed at the object
    \item \textbf{Control Glance:} Users look again at the object to maintain the pointing position; then the pointing gesture ends and they move their arms back
\end{itemize}
They also observed different behaviour for different users during the last control glance phase, wherein for 57\% of the gestures, users looked again (making two control glances) to further maintain the pointing position, while other users did not take the control glance at all in 8\% of the gestures. Moreover, they reported that free-hand pointing does not increase the cognitive load of the driver in terms of constant driving speed during gestures performance, which was later confirmed by Roider et al.~\cite{Roider2017} as well. 

In conclusion, previous multimodal referencing methods mostly focused on in-car interaction, unlike this work, which focuses on interaction with the environment outside the car.
Additionally, both in-car and outside-the-car existing methods either focused on a stationary car scenario where performing the referencing gestures was the primary task, or they had a short driving route, unlike our approach, which focuses on driving as the primary task and performing the referencing gestures as the secondary task in a long driving route.

\section{Method}

\begin{figure}[!t]
	\begin{center}
		\includegraphics[width=0.85\linewidth]{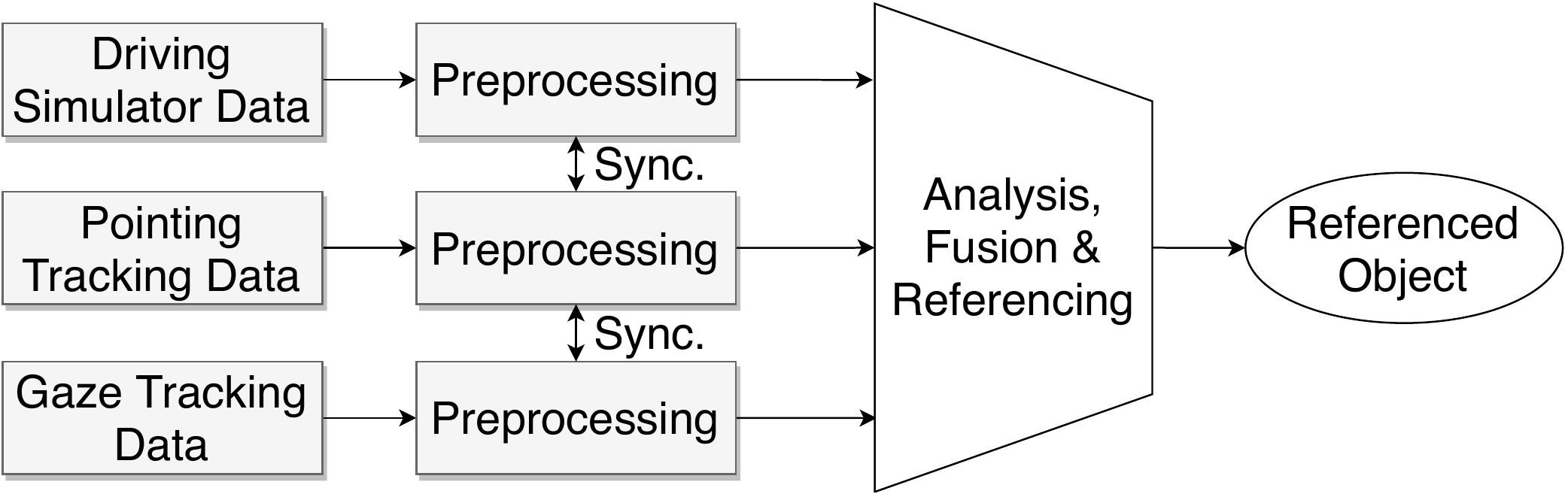}
	\end{center}
	\caption{Overview of system architecture}
	\label{fig:implementationarchitecture}
\end{figure}

A within-subject counterbalanced experiment was designed in a medium fidelity driving simulator~\cite{math2013opends}. We chose a driving simulator instead of a real car scenario for better control over the study and safety aspects. A driving simulator might have some influence on participant's behaviour in comparison to a real car scenario as it is a more relaxed environment. However, we hypothesize that drivers tend to reference objects outside a moving vehicle only in relaxed and easy driving situations (i.e. drivers prioritize the primary task of driving~\cite{rumelin2013free}). 
Therefore, a simulated driving scenario matches the real experience to a great extent for this task.
\autoref{fig:implementationarchitecture} shows an architecture overview of the desired system. Users' pointing and gaze along with their driving route were tracked for the referencing task. Each of the data channels was processed separately to reach a common data format (i.e. common coordinate system) among them while synchronising their internal clocks. Finally, modalities' interactions were investigated and fusion approaches were attempted to identify the referenced objects.

\subsection{Design and Procedure}

\begin{table}[b]
\centering
\caption{Driving conditions}
\resizebox{0.75\linewidth}{!}{\begin{tabular}{c c c c} 
\hline\hline
Condition & Autonomous Driving             & Distance & Density   \\ 
\hline
1        & No              & Near      & Dense     \\
2       & No               & Far     & Dense     \\
3       & No               & Near      & Non-Dense      \\
4       & No               & Far     & Non-Dense      \\
5  & Yes & Far     & Dense      \\
\hline
\end{tabular}}
\label{table:conditions}
\end{table}

The experiment consisted of a driving task in a driving simulator. A pilot study was conducted before the main experiment to enhance the design and ensure feasibility. Each participant drove for 40 minutes on the right of a two-lane road with no traffic at a maximum speed of 60 km/hour while performing a secondary task of referencing (i.e. pointing and looking at) objects (i.e. buildings) 
which will be called \textit{Point Of Interest} (PoI) hereafter. 
Participants were informed which PoI they had to point at in real time by displaying the targets on a small tablet and signalled with an auditory cue.

The driving route was designed as a star shape containing five corners (i.e. sections). To ensure that there is no confounding influence of the angled parts, target buildings and distractors were only presented while driving straight. Each corner of the star corresponded to different conditions with respect to environmental parameters and driving mode. The environmental parameters were the PoI distance from the road and number of distractors in the environment as seen in~\autoref{table:conditions}. The first and second condition had near and far PoIs respectively in a dense environment (i.e. many distractors around the PoI) while the environment was less dense in the third and fourth condition (i.e. few to no distractors around the PoI). The last condition was the same as the second in terms of density and distance, but the vehicle drove autonomously. 

An online \textit{pre-study} was conducted to determine the optimal values for distance and density levels. Another goal for the pre-study was to choose the visual appearance of the PoIs (i.e. their shape and color), to determine the salience of the PoI (i.e. target) against the distractor buildings. A medium-salient PoI is required to keep moderate discoverability, to avoid confounding factors \sugg{(i.e hard enough not to spot with peripheral view but easy enough to find it eventually)}. Seventeen participants (59\% male) with a mean age of 31.12 years (SD = 15.47) completed this online pre-study.

Participants performed the referencing task 24 times per condition in a counterbalanced manner. Two-thirds of the PoIs were located on the right side of the road, since the hand tracking only captures right-handed pointing, which could affect pointing at left-oriented PoIs (i.e. PoIs located on the left side of the road). Between two consecutive PoI notifications, there was a time gap of 10 to 20 seconds, giving participants enough time for visual search and referencing while maintaining the primary driving task. The road had no traffic to accommodate this relatively small time gap.

\subsection{Apparatus}

\begin{figure}[t]
	\begin{center}
		\includegraphics[width=0.9\linewidth]{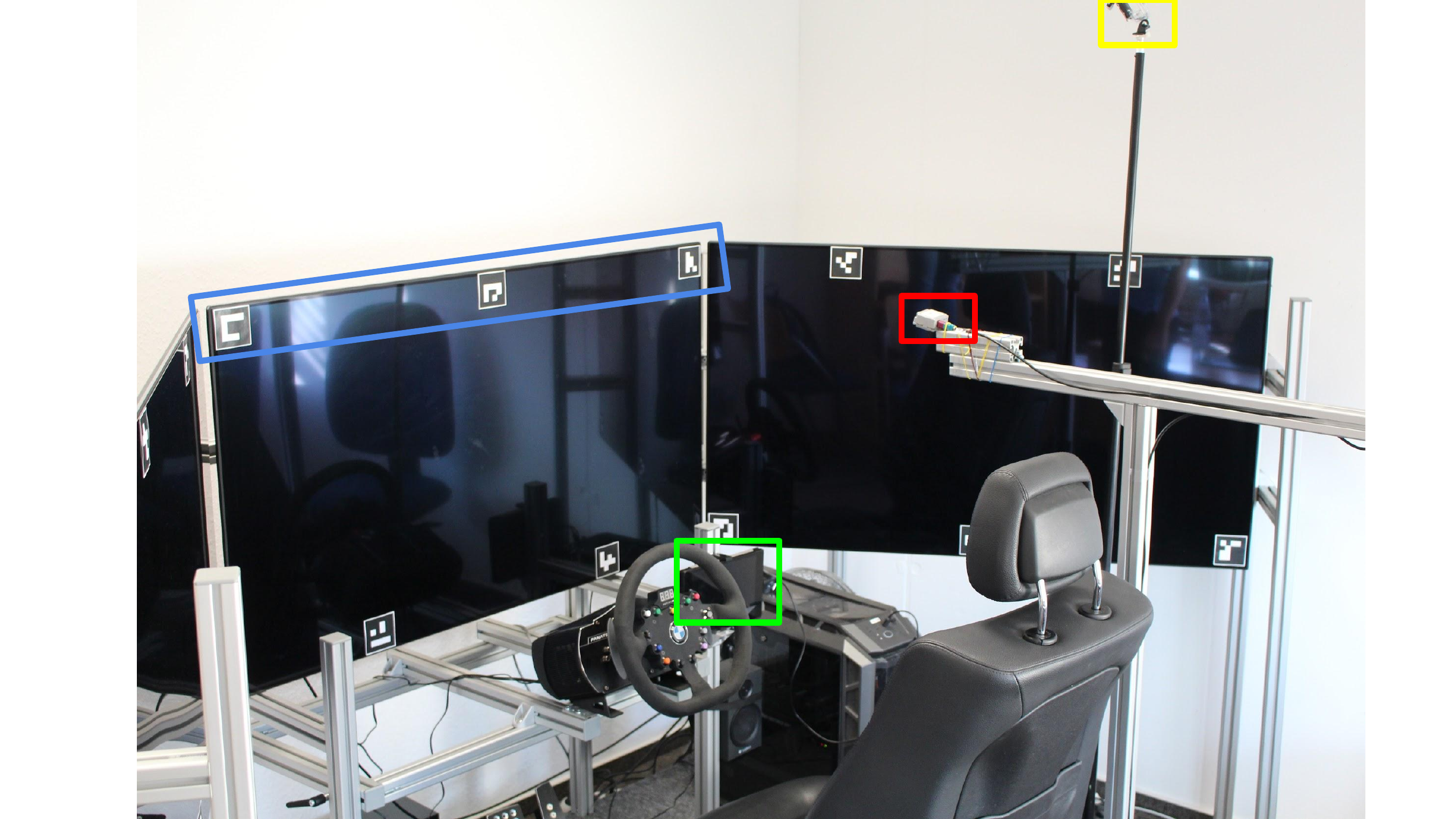}
	\end{center}
	\caption{Driving simulator setup overview. The PoI notification tablet is highlighted in green, the hand pointing camera in red, the GoPro camera in yellow and a sample of the ArUco markers highlighted in blue.}
	\label{fig:setup2}
\end{figure}

The driving simulator (\autoref{fig:setup2}) was situated in an enclosed room to ensure minimum disturbance. It consisted of three 55-inch LCD screens with a steering wheel, pedals and two left and right speakers. ArUco~\cite{Aruco1,Aruco2} markers were attached to fixed known locations at the edges of the LCD screens which were used in eye gaze vector mapping. The PoI image was presented on a tablet next to the steering wheel and was preceded by a sound notification. The hand tracking camera and the experiment recording camera (a GoPro camera) were situated to the right of the participant. The hand tracking camera was a state-of-the-art non-commercial prototype especially designed for in-vehicle control. The eye gaze tracker used was a pair of \textit{SMI Eye Tracking Glasses}\footnote{\url{https://imotions.com/hardware/smi-eye-tracking-glasses/}}.

\subsection{Participants}

In total, 73 participants were recruited for the study. However, 34 participants were excluded for the following reasons: technical problems in pointing and gaze trackers such as severe frame drops and failure to save data (30 participants), premature termination due to motion sickness (2 participants) or improper task execution (2 participants). The data of the remaining 39 participants were manually verified to ensure correct synchronization with no technical problems. The remaining participants (46\% male) with a mean age of 25.87 years (SD = 6.26) completed the entire driving route while referencing objects using their right hand as instructed, to stay within the pointing tracker range. 83\% of the participants were right-handed. 14\% were left-handed and 3\% reported ambidexterity.

\subsection{Coordinate System and Features Extraction}

\begin{figure}[b]
	\begin{center}
		\includegraphics[width=0.45\linewidth]{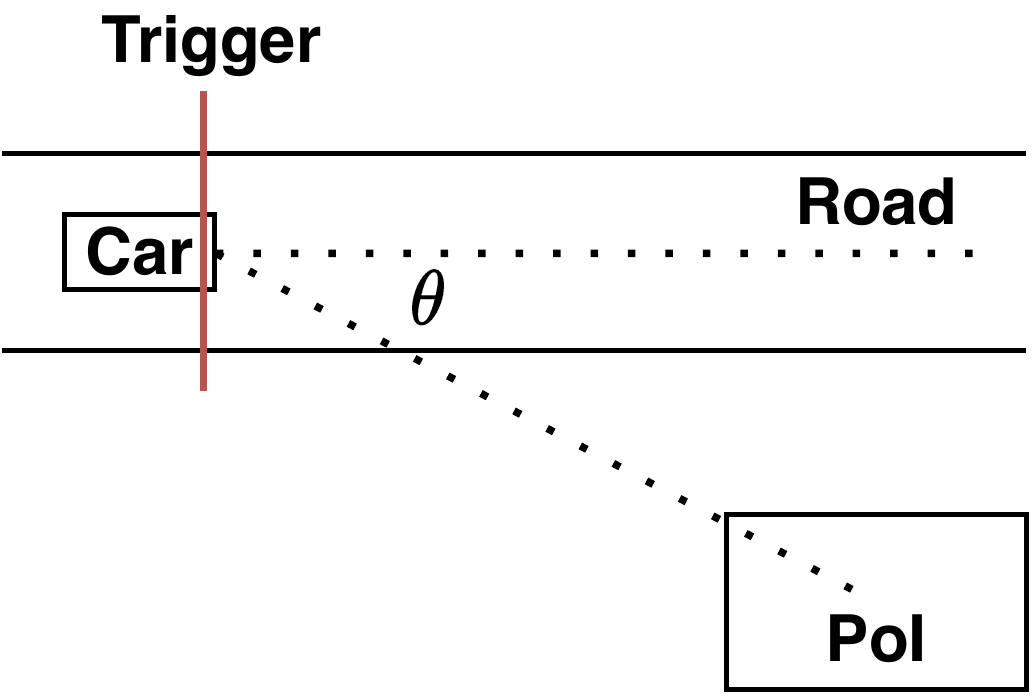}
	\end{center}
	\caption{An example of the angular coordinate value at a given time instance.}
	\label{fig:coordinatesystem}
\end{figure}

\begin{figure}[t]
	\begin{center}
		\includegraphics[width=\linewidth]{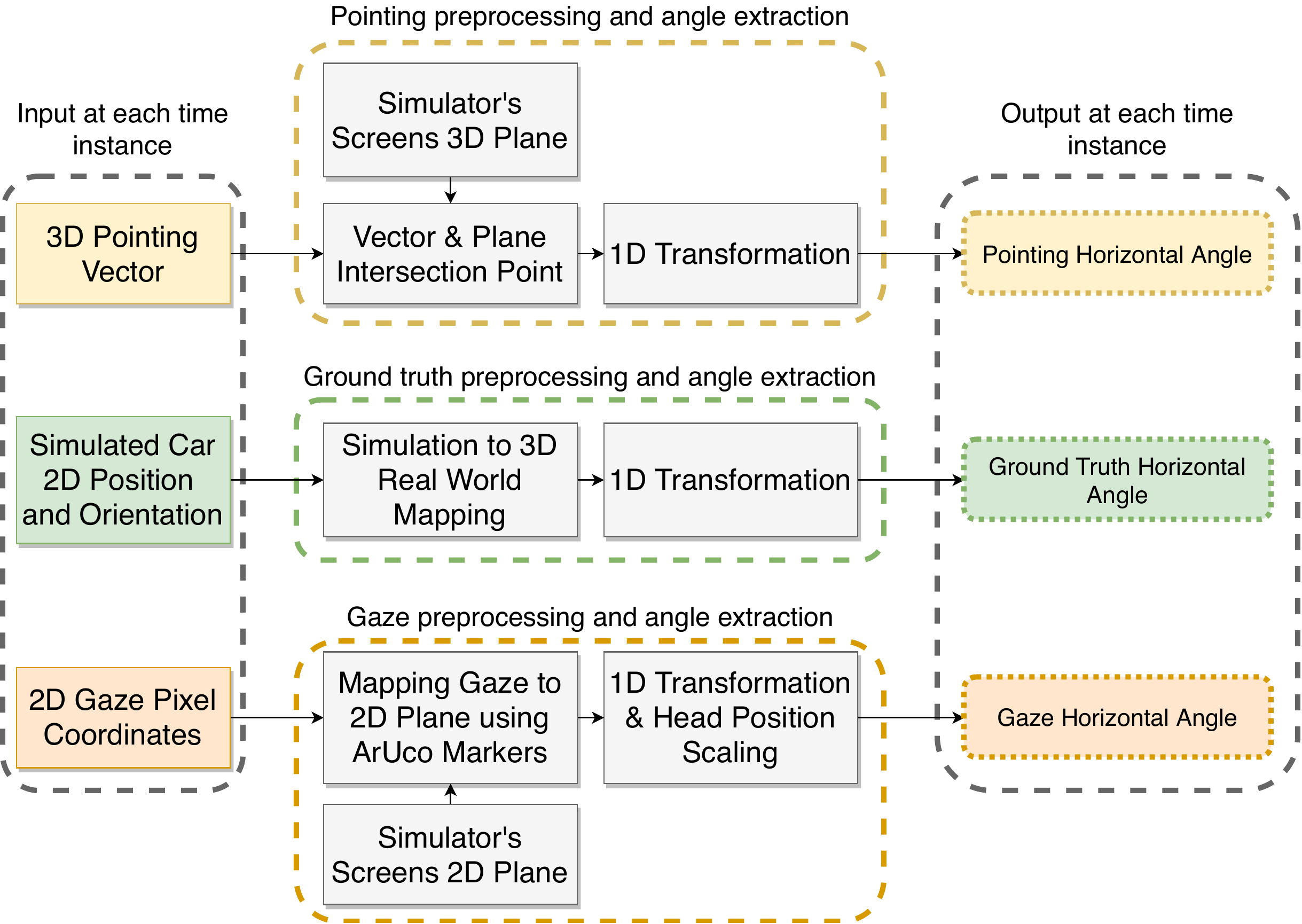}
	\end{center}
	\caption{Overview of our approach showing the analysis process for each system and highlighting the input (raw data) and output (horizontal angles) formats.}
	\label{fig:preprocess_angle_extract}
\end{figure}

Each of the pointing and gaze tracking systems had their own coordinate systems which needed to be mapped to the simulation environment. We only considered horizontal angle in our approach, similar to Kang et al.~\cite{kang2015you}, as there is no overlap in the PoI vertical position. Thus, we used a 1D cylindrical coordinate system as a common coordinate system (see~\autoref{fig:implementationoverview}).~\autoref{fig:coordinatesystem} shows an example of the coordinate system at a given time instance where $\theta$ is the angle between the vehicle's centre line and the line connecting the vehicle with the PoI. It was considered as the ground truth (GT) angle which is calculated at all time frames. However, this GT angle was further relaxed with the addition of the PoI's angle span (i.e. the relative angle corresponding to building's width) since users could point at the edge of the PoI instead of the centre. ~\autoref{fig:preprocess_angle_extract} illustrates the process of mapping the tracking systems to the simulation system and extracting the 1D horizontal angles.


\paragraph{Ground Truth Preprocessing and Angle Extraction}

For each time instance, the GT angle ($\theta$ in~\autoref{fig:coordinatesystem}) was calculated in multiple stages as seen in~\autoref{fig:preprocess_angle_extract}. First, the car's virtual 2D location and orientation along with the PoI's virtual 2D location were mapped into the world 3D Cartesian coordinate system through linear transformation. Next, the 3D coordinate system was transformed into a 1D cylindrical coordinate system where the horizontal ground truth was extracted with a 0.1 degree resolution. Since the user was allowed to point anywhere at the PoI building, the building width was added as a margin to the ground truth angle in further calculations. 
\sugg{The advantage of the previously mentioned approach is that it is only time-dependent and not speed-or distance-dependent, because it takes the car location information from each participant data file (and not from a hard-coded driving path which is usually the case for GT calculation). This allowed for a personalized analysis of the data, as each participant had his own GT based on his driving speed and behaviour.}

\paragraph{Gaze Preprocessing and Angle Extraction}

The gaze tracker outputs gaze pixel location inside the recorded surroundings. These 2D gaze pixel (x,y) coordinates were mapped to a reference coordinate system using translation transformation as seen in~\autoref{eqn:mapping}:
\begin{equation}
\label{eqn:mapping}
\begin{split}
(x,y)_{gaze\:new} & = (x,y)_{gaze\:image} \\-  & [(x,y)_{ArUco\:image}-(x,y)_{ArUco\:reference}] 
\end{split}
\end{equation}
where $(x,y)_{gaze\:new}$ is the gaze coordinates in the new common coordinate system, $(x,y)_{gaze\:image}$ is the one calculated from the frames and $(x,y)_{ArUco\:image}$, $(x,y)_{ArUco\:reference}$ are the ArUco marker's coordinates calculated from the frames and set in the common coordinate system, respectively.
The horizontal angle was calculated by linear transformation from x-coordinates (ranging from -1280 to 1280) to angle coordinates (ranging from -90 to 90). However, this linear transformation had a scale of one for the middle screen only (range from -45 to 45) and scale of half for the left and right screens (range from -90 to -45 and 45 to 90, respectively). This is due to the 45 degree inclination of the side screens (see ~\autoref{fig:implementationoverview}).
The final horizontal angle was scaled per participant to adjust for different seating positions. Average x-coordinate difference between two predefined ArUco markers was calculated per participant and divided by the x-coordinate difference between the same two markers in the common coordinate system; then, the resultant was multiplied with the previously calculated horizontal angle as seen in~\autoref{eqn:mapping_scaling}: 
\begin{equation}
\label{eqn:mapping_scaling}
\theta_{new} = \theta_{old} * \frac{(x_{ArUco1} - x_{ArUco2})_{participant}}{(x_{ArUco1} - x_{ArUco2})_{reference}}
\end{equation}
        
The final gaze angle had a resolution of 0.01 degrees; however, there was an error margin of $\pm 3$ degrees due to rounding approximations.

\paragraph{Pointing Preprocessing and Angle Extraction}

For each time instance, the hand tracker gave the position of the fingertip and the vector of pointing in the 3D real world coordinate system referenced to the tracker position. The tracker also classified the hand gesture type (i.e. pointing, sliding right, sliding left, etc.). This classification was used to mask only pointing gestures and passed the fingertip position and the pointing vector as input to next stages (as seen in~\autoref{fig:preprocess_angle_extract}). To calculate the angle of pointing, the LCD screens' 3D planes were calculated with respect to the camera, then intersected with the pointing vector to get the intersection point coordinates in the 3D world Cartesian system with respect to the participants (instead of the camera). Similar to gaze and GT, the 3D Cartesian coordinates were transformed into 1D cylindrical coordinates and the final horizontal pointing angle was obtained. It had a resolution of 1 degree and an error margin of $\pm 4$ degrees.

\subsection{Performance Metrics}

\begin{figure}[b]
	\begin{center}
		\includegraphics[width=\linewidth]{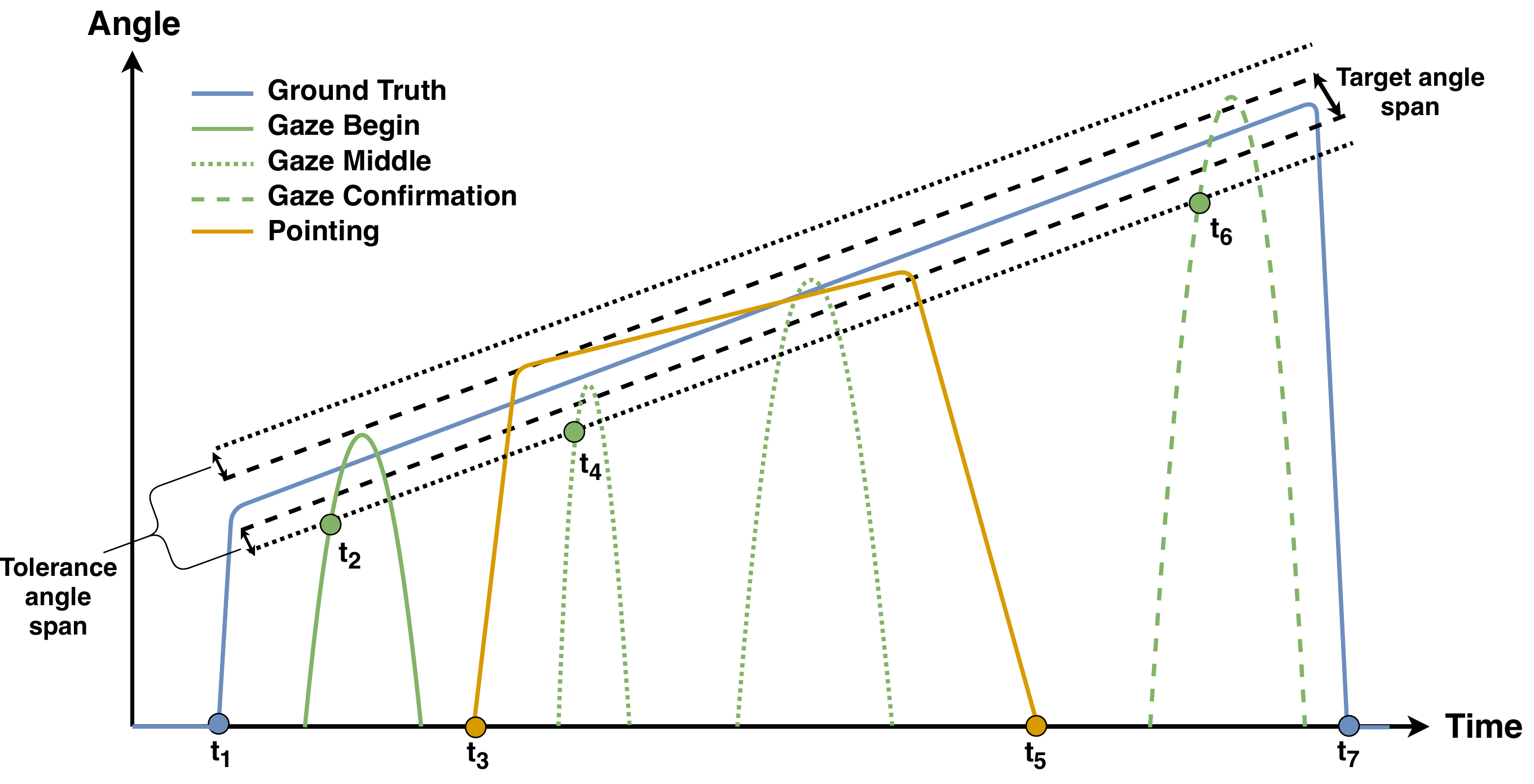}
	\end{center}
	\caption{A representation of our behavioural model for pointing and gaze angles vs. the ground truth \textbf{for a single trigger} with (simplified) horizontal angle on the y-axis and time on the x-axis. Where $t_{1}$ and $t_{7}$ are the start and end times of the ground truth calculation (i.e. start of PoI target being displayed on the tablet).  $t_{2}$ is the start time of the first gaze at the target PoI. $t_{3}$ and $t_{5}$ are the start and end times of the pointing gesture, respectively. $t_{4}$ is the start time of the first gaze that occurs after the start of the pointing gesture and $t_{6}$ is the start time of the first gaze that occurs after the end of the pointing gesture.}
	\label{fig:performancemeasurement}
    \end{figure}

To study the referencing behaviour systematically, a behavioural model was constructed from the related work and observations on our task of referencing predetermined PoIs while driving.~\autoref{fig:performancemeasurement} shows a simplified representation of this model for a right-oriented PoI. The predetermined PoIs appeared at a certain angle which slowly increased until the PoI disappeared to the far right. Participants gazed at it several times to compare it to the given image of the PoI and to confirm accurate referencing. Users pointed at the PoI only once as instructed. The same model represents left-oriented PoI but with negatively increasing angle values.
Task performance was assessed by analysing the effect of several independent variables on multiple dependent ones which were used in related work for assessing multimodal gesture recognition systems. 

\subsubsection{Independent Variables (IVs)} The previously mentioned conditions are the main independent variable. They were further subdivided to several IVs based on distance, density and PoI orientation. Furthermore, the personalized behaviour of participants was considered among these conditions.

\subsubsection{Dependent Variables (DVs)}

The dependent variables are divided into two main categories as follows. They were calculated as a mean value per independent variable. For example, pointing accuracy is calculated per participant, density, distance and PoI orientation for each participant.

\paragraph{Performance-related DVs (i.e. Accuracy)}

    For each trigger, the pointing gesture was considered accurate if the pointing angle was in the range of the PoI angular width plus a fixed tolerance angle span of 10 degrees for 400 milliseconds or more~\cite{mcleod1987visual}. 
    \textit{Pointing accuracy DV} was then calculated per independent variable by dividing the sum of \textit{accurate pointing} triggers by the sum of \textit{pointed at} triggers (as seen in~\autoref{eqn:pointingaccuracy}). For example, condition one for each participant contained 24 triggers; a participant only pointed at 21 triggers (due to losing focus or similar issues); if he accurately pointed at only 17 triggers, then the accuracy of pointing in condition one for this participant was 17 divided by 21.
    \begin{equation}
        \label{eqn:pointingaccuracy}
        (Pointing\_Accuracy)_{per\,IV} = (\frac{\sum Accurate\_Pointing}{\sum Pointing})_{per\,IV}
    \end{equation}
    
    Similarly, \textit{gaze accuracy DV} is calculated as the sum of \textit{accurate gaze} triggers divided by the sum of \textit{pointed at} triggers (as seen in~\autoref{eqn:gazeaccuracy}). However, \textit{accurate gaze} had an extra time condition where gaze should only be considered during the pointing time window with a tolerance of 500 milliseconds before and after (i.e. \textit{accurate gaze} is only considered in the time from $(t_{3} - 500ms)$ to $(t_{5} + 500ms)$ in~\autoref{fig:performancemeasurement}). \textit{Accurate gaze} is also considered if all previous conditions are met for 400 milliseconds or more~\cite{mcleod1987visual}. 
    \begin{equation}
        \label{eqn:gazeaccuracy}
        (Gaze\_Accuracy)_{per\,IV} = (\frac{\sum Accurate\_Gaze}{\sum Pointing})_{per\,IV}
    \end{equation}

\paragraph{Timing-related DVs} Four dependent variables were calculated from our model as follows: 
\begin{itemize}
        \item \textit{Detection Time} describes the time that passed between the presentation of the target building on the tablet and the time when the building was first gazed at (discovered) by the participant ($t_{2} - t_{1}$ in~\autoref{fig:performancemeasurement}).
        \item \textit{Pointing Time (Reaction Time)} describes the time interval between the presentation of the building on the tablet and the onset of the pointing gesture towards the PoI ($t_{3} - t_{1}$ in~\autoref{fig:performancemeasurement}). It could also be described as reaction time (i.e. the time it takes for the participant to react to a given PoI).
        \item \textit{Modality Delta Time (Action Time)} is the time interval between the onset of the gaze on the PoI and the onset of the pointing gesture ($t_{3} - t_{2}$ in~\autoref{fig:performancemeasurement}). This showed whether pointing gesture usually follows the gaze or vice versa and how much participants wait to start pointing. If the first gaze occurred after the pointing, this value would be negative.
        \item \textit{Confirmation Time} describes any confirmation gazes that the participant made back at the PoI to confirm if it had truly been the correct building ($t_{6} - t_{5}$ in~\autoref{fig:performancemeasurement}).
    \end{itemize}
Additionally, dependent variables relating to duration and frequency of pointing and gaze gestures were calculated as follows:
\begin{itemize}
    \item \textit{Pointing Duration} was calculated for each trigger ($t_{5} - t_{3}$ in~\autoref{fig:performancemeasurement}) and averaged per independent variable.
    \item \textit{Gaze Duration}: three aspects of gazing behaviour were identified in the analysis. During a visual inspection of the data and similar to~\cite{rumelin2013free}, it became apparent that gazes towards a PoI could be divided into three categories (as seen in green in~\autoref{fig:performancemeasurement}) as follows:
    \begin{enumerate}
        \item \textit{Gaze Begin} describes gazes towards the PoI before the pointing gesture.
        \item \textit{Gaze Middle} describes gazes towards the PoI during the pointing gesture. 
        \item \textit{Gaze Confirmation} is identified as one or several gazes back towards the PoI, after the pointing has ended (possibly to confirm the correctness of the targeted PoI).
    \end{enumerate}

    \item \textit{Gaze Frequency}: Similarly, the frequency of occurrence of gaze was calculated separately for each gaze gesture type. As an example in~\autoref{fig:performancemeasurement}, \textit{Gaze Begin Frequency} is one, \textit{Gaze Middle Frequency} is two and \textit{Gaze Confirmation Frequency} is one as well.

\end{itemize}

\subsection{Auxiliary Hypotheses}

Five auxiliary hypotheses were developed specifically for the performance metrics for interpretation based on literature findings. Better performance in these hypotheses means higher pointing and gaze accuracy and lower reaction time:

\begin{itemize}
    \item \textbf{Hypothesis 1}: A high density of buildings in the scenery leads to a higher mental load, since there are more stimuli that require processing and comparison. Therefore, performance should be better in the non-dense condition.  
    \item \textbf{Hypothesis 2}: A high distance between the target building and the road 
    leads to lower detection performance. Therefore, performance should be better in the near condition.
    \item \textbf{Hypothesis 3}: Overall performance should be better in the autonomous driving than in the normal driving condition, since participants have to perform two tasks when driving normally, which is more difficult. 
    \item \textbf{Hypothesis 4}: Performance should be better for targets on the right side of the road. There were more PoIs on the right side, which could lead to a higher learning effect. 
    \item \textbf{Hypothesis 5}: During a multimodal referencing task, accuracy of gaze should be better than accuracy of pointing.
\end{itemize}

\section{Results}

\begin{table}[b]
\centering
\caption{Descriptive statistics and Intraclass correlation result per DV}
\resizebox{\linewidth}{!}{\begin{tabular}{c c c} 
\hline \hline

Dependent Variable                                   & Mean (Std Dev.) & ICC in percentage  \\ 
\hline
Pointing Accuracy                  & 68.97\% (17.15)     &     -    \\
Gaze Accuracy                         & 83.48\% (10.22)   &     -     \\
Detection Time                          & 1.81 sec. (0.48)     & 3.5\%         \\
Pointing Time                          & 3.69 sec. (2.58)    & 53.7\%         \\
Modality Delta Time                     & 1.91 sec. (2.54)  & 38.4\%      \\
Confirmation Time                       & 2.43 sec. (0.44)  & 1.6\%      \\
Pointing Duration                       & 1.88 sec. (1.09)  & 41.5 \%      \\
Total Gaze Duration                     & 5.04 sec. (0.72)  & 4.7\%          \\
Gaze Begin Duration                     & 1.83 sec. (0.99)  & 21.7\%          \\
Gaze Middle Duration                    & 0.67 sec. (0.41)  & 12.7\%  \\
Gaze Confirmation Duration              & 2.54 sec. (0.72)  & 7.8\%       \\
\hline
\end{tabular}}
\label{table:descrstatistics}
\end{table}

The results section is divided into two main parts. The first part is the statistical analysis of the data, while the second part explains the results of several clustering approaches.

\subsection{Statistical Results}

In this section, results from the descriptive statistics, several intraclass correlation (ICC)~\cite{Johnson2011} analyses and inferential statistical analyses are described.

\subsubsection{Descriptive statistics}

\autoref{table:descrstatistics} shows the means and standard deviations for all dependent variables. Average gaze accuracy was better than average pointing accuracy during the multimodal interaction. Average pointing time (3.69 seconds) was relatively short compared to the time gap (10 to 20 seconds) between two consecutive PoI appearance which suggests swift pointing behaviour.
Further investigation shows that the average GT angle at the middle of the pointing frame was found to be $\pm 9.8$ degrees (i.e. 9.8 degrees on the right or the left) which confirms this finding. 
Moreover, average duration of gaze during pointing (0.67 seconds) was quite short compared to the average pointing duration (1.88 seconds) which suggests that participants don't keep their eyes on the PoI during the entire pointing interaction duration.
\autoref{table:freqstatistics} shows the average frequency of gazes for the three defined gaze types per participant and sum frequency of gazes per trigger. 

\begin{table}[t]
\centering
\caption{Average gaze frequency percentage per participants and number of triggers for each gaze category}
\setlength{\tabcolsep}{3pt}

\renewcommand{\arraystretch}{1.25}
\resizebox{\linewidth}{!}{\begin{tabular}{c|cc|cc|cc}
\hline\hline
\multirow{2}{*}{Frequency~}           & \multicolumn{2}{|c}{Gaze Begin}      & \multicolumn{2}{|c}{Gaze Middle}     & \multicolumn{2}{|c}{Gaze Confirmation}  \\
\cline{2-7}
& \% Partic. & \% Trig. & \% Partic. & \% Trig. & \% Partic. & \% Trig.       \\
\hline
0          & -               & 25.2\%           & 46.0\%              & 52.1\%                   & -             & 17.0\%                       \\
1          & 56.8\%         & 35.7\%           & 51.3\%              & 35.9\%                   & 13.5\%       & 18.3\%                       \\
2          & 27.0\%         & 20.7\%           & 2.7\%                & 8.0\%                   & 27.0\%        & 18.0\%                       \\
3          & 13.5\%         & 8.2\%            & -                    & 2.4\%                   & 51.4\%        & 16.7\%                      \\
4          & 2.7\%           & 4.9\%            & -                    & 1.0\%                   & 8.1\%         & 12.6\%                       \\
$\geq 5$  & -               & 5.3\%            & -                     & 0.6\%                   & -              & 17.4\%                       \\
\hline
\end{tabular}}
\label{table:freqstatistics}
\end{table}

\subsubsection{Intraclass correlations results}

\autoref{table:descrstatistics} also shows intraclass correlations (ICCs) for all timing-related dependent variables. The average ICC across all variables was 20.62\% which confirms the person-specific behaviour of participants during the referencing task~\cite{brown2016exploring,brown2014performance,rumelin2013free}.

\subsubsection{Statistical inference}

Several multivariate analyses of variance~\cite{smith1962multivariate} (MANOVA) were chosen for its suitability to assess the auxiliary hypotheses for multiple dependent variables. Statistical significance is compared against an alpha level of 5\% (i.e. p-value < 0.05).
The statistical preconditions for a MANOVA were met. Performance was measured by \textit{pointing accuracy}, \textit{gaze accuracy} and \textit{pointing time} dependent variables. 
\paragraph{Distance and density} A 2x2 within-subject MANOVA was conducted to measure the effects of density and distance on the defined performance DVs. There was a significant effect in performance for density (Pillai trace=.72, F(3,32)=27.62, p<.001, $\eta_p^2$=.72). This difference is significant for all three performance variables (pointing accuracy: F(1,34)=34.94, p<.001, $\eta_p^2$=.51; gaze accuracy: F(1,34)=52.79, p<.001, $\eta_p^2$=.60; pointing time: F(1,34)=56.31, p<.001, $\eta_p^2$=.62). Pointing and gaze accuracy were higher and pointing time was faster for the non-dense conditions.
There was no significant overall effect on performance measurements for distance levels (p=.254) and no significant interaction of distance and density (p=.155).
\paragraph{Autonomous vs. normal driving} 
A within-subject MANOVA was conducted to compare conditions two and five. There was a significant difference between the conditions for the defined performance variables (Pillai trace=.49, F(3,32)=10.91, p<.001, $\eta_p^2$=.49). However, this difference is only significant for gaze accuracy (F(1,34)=21.46, p<.001, $\eta_p^2$=.39) where it was higher in the autonomous drive condition (i.e. condition five). There was no significant difference for pointing accuracy (p=.456) and pointing time (p=.116).
\paragraph{Left vs. right PoI orientation} A within-subject MANOVA was conducted to compare left and right PoI orientation for the defined performance variables. There was a significant difference in performance for PoI orientation (Pillai trace=.54, F(3,32)=12.48, p<.001, $\eta_p^2$=.54). This difference can be found significantly in gaze accuracy (F(1,34)=16.72, p<.001, $\eta_p^2$=.33) and pointing time (F(1,34)=8.61, p=.006, $\eta_p^2$=.20). PoI oriented on the right side were pointed at faster and gazed at more accurately. There was no significant difference for pointing accuracy (p=.965). 
\paragraph{Pointing vs. gaze accuracy} Lastly, a within-subject MANOVA was conducted to compare pointing and gaze accuracy. Gaze accuracy was significantly higher compared to pointing accuracy (Pillai trace=.42, F(1,37)=26.77, p<.001, $\eta_p^2$=.42).

\subsection{Clustering Results}

Cluster analysis was used to find patterns in participants' behaviour beyond their individuality for possible fusion approaches that fits these clusters. In this section, clustering the participants based on their behaviour was attempted, using the previously mentioned metrics as features.

Two clustering approaches were attempted: k-means non-hierarchical clustering and agglomerative hierarchical clustering. Both approaches yielded similar results. Therefore, only the results of k-means clustering are presented here. Clustering was done using participants' performance-related dependent variables and time-related dependent variables as features. An elbow curve was used to determine the best value for \textit{k}.

\subsubsection{Clustering using \textit{performance-related} dependent variables}

            \begin{figure}[b]
	\begin{center}
		\includegraphics[width=0.45\linewidth]{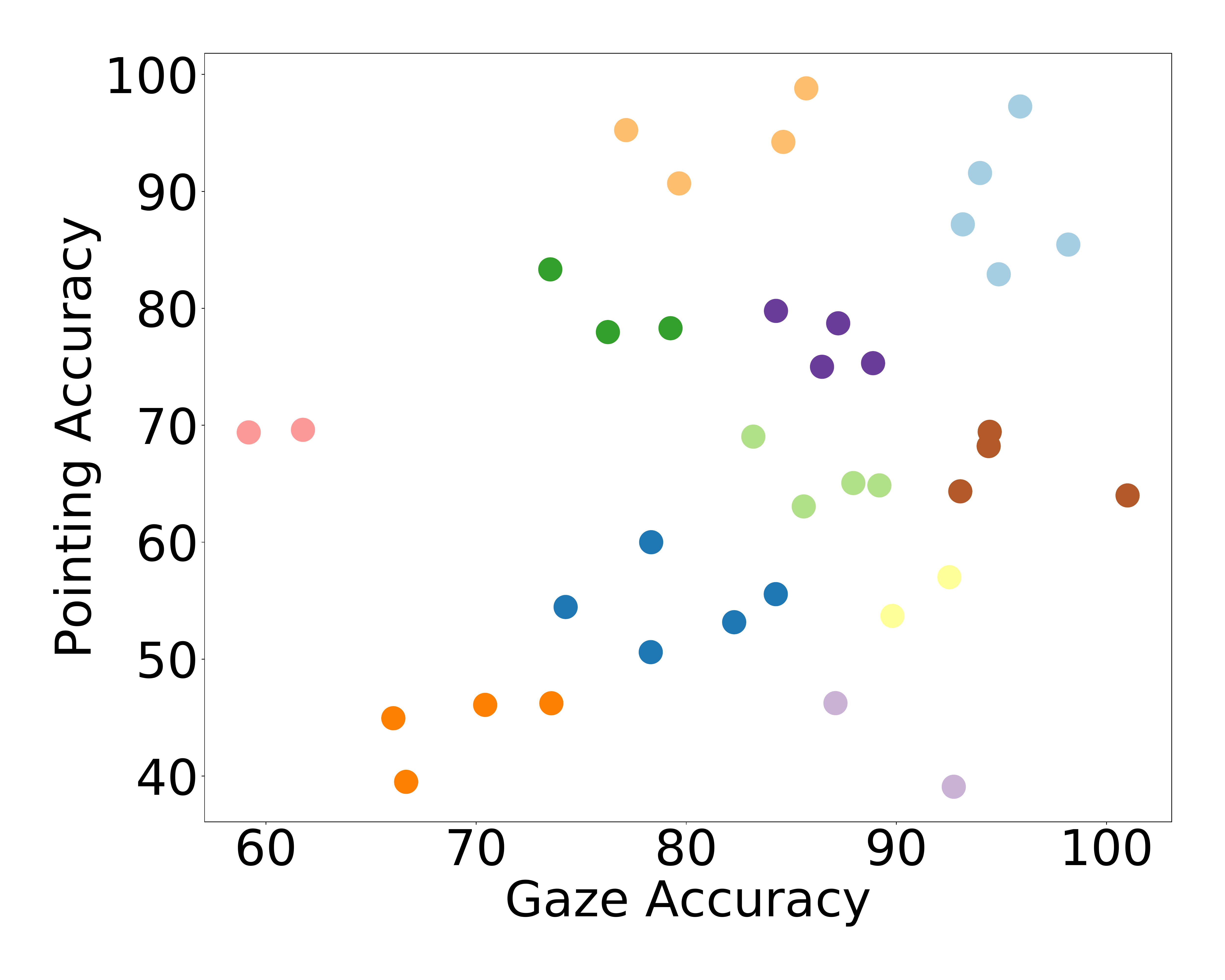}
		\includegraphics[width=0.45\linewidth]{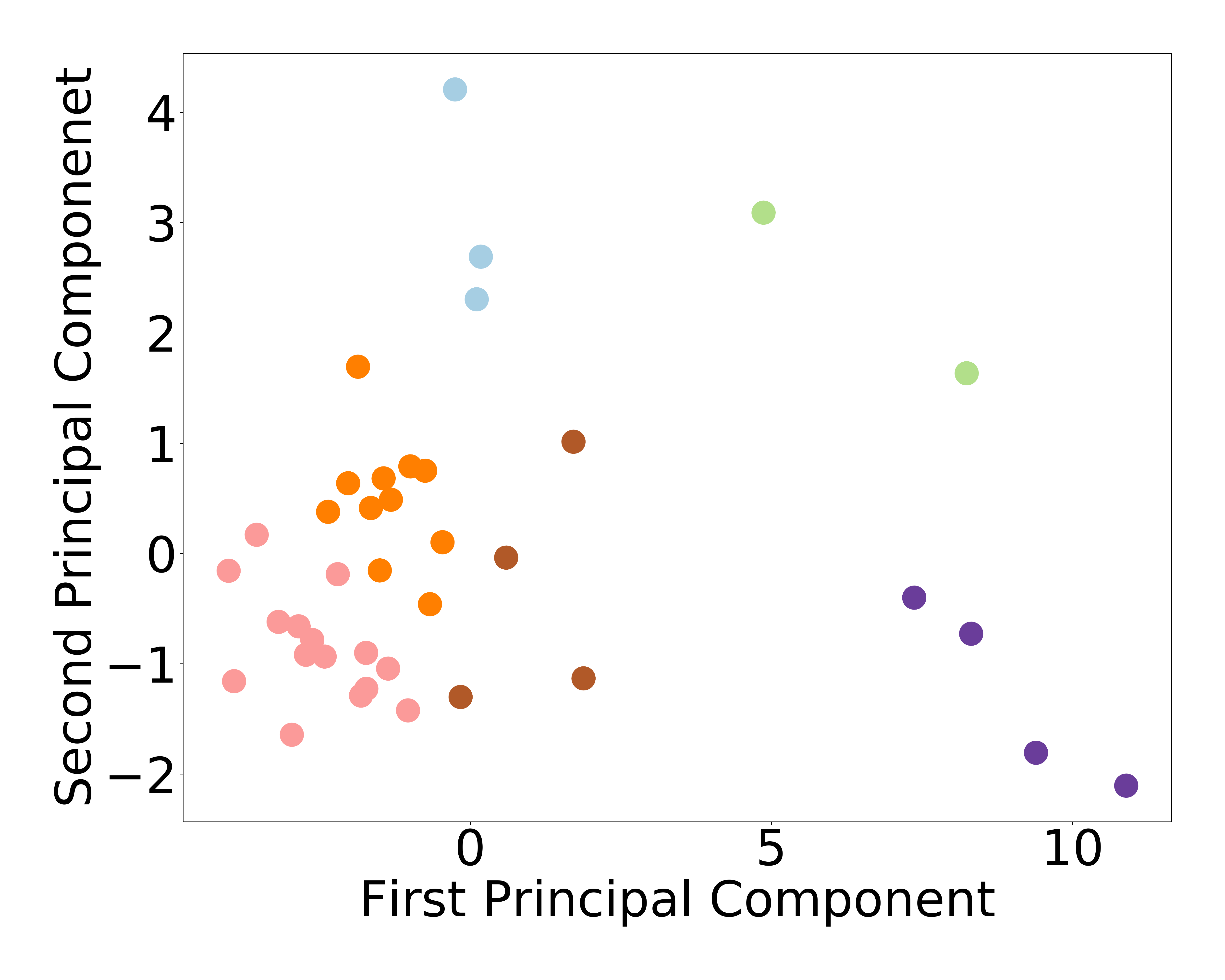}
	\end{center}
	\caption{K-means clustering using performance-related features (left, with K=11) and timing-related features (right, with K=6)}
	\label{fig:kmeansclusteraccuracy}
    \end{figure}

\begin{figure}[t]
	\begin{center}
		\includegraphics[width=0.75\linewidth]{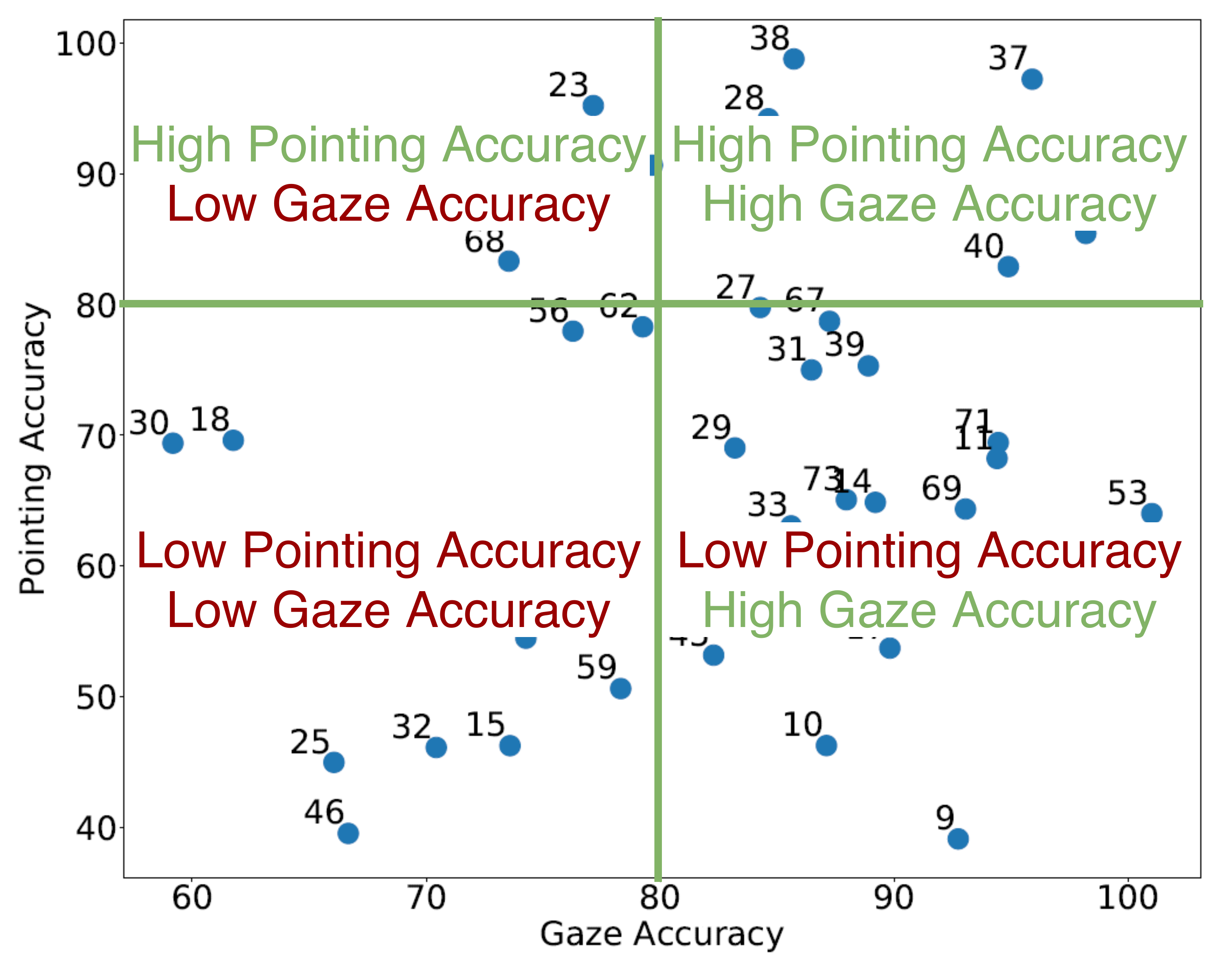}
	\end{center}
	\caption{Participants' distribution across pointing accuracy (y-axis) and gaze accuracy (x-axis) with four quadrants clustering at a threshold of 80\%}
	\label{fig:distraccuracywithquadrants}
    \end{figure}

\autoref{fig:kmeansclusteraccuracy} shows k-means clustering output using gaze and pointing accuracy as features. 
To get rid of specific condition influence on the total average value, clustering pointing and gaze accuracy per each independent variable was investigated; all analyses showed similar results to the statistical inference analysis.
Typical clustering algorithms did not yield meaningful clusters since both pointing accuracy and gaze accuracy dependent variables are normally distributed with similar standard deviations and only shifted means. Instead, heuristic clustering was done on the participant's distribution by dividing the participants into four quadrants in terms of accuracy (see~\autoref{fig:distraccuracywithquadrants}), which could be used for modality switching based on tracking performance.

\subsubsection{Clustering using \textit{time-related} dependent variables}

Similar to clustering using performance-related dependent variables, k-means clustering was done using all time-related dependent variables as features. To visualize the clustering output, principal component analysis (PCA) was applied to the 11D features to reduce them to 2D.~\autoref{fig:kmeansclusteraccuracy} shows the clustering output. 

\section{Discussion and Limitations}
This section is divided into two parts. The first part describes our interpretation of the results and how can it be utilized in multimodal referencing, while the second part explains the limitations of our approach. 

\subsection{Discussion}

\subsubsection{Descriptive statistics}
Descriptive statistics shows that on average, gaze accuracy
was better than pointing accuracy during the multimodal interaction. However, as previously mentioned, the gaze accuracy calculation already included pointing modality timing so it was not a stand-alone modality. It also shows that pointing time was relatively short, which means that users did not wait for the PoI to come closer before pointing, but rather pointed quickly at distant PoIs. The average ground truth angle confirms this finding as it was $\pm 9.8$ degrees, while a close PoI ground truth angle was in the range of 30 to 40 (right-oriented PoI) or -30 to -40 (left-oriented PoI) degrees. 

Regarding gaze type frequency analyses, `gaze begin' shows that half of the participants looked at the target (i.e. PoI) two to four times before pointing, while half of the participants looked only once. For `gaze middle', almost all participants did not look at the target while pointing, or looked only once, which suggests that users don't get visually distracted by the referencing task during driving. For `gaze confirmation', all participants looked at the target at least once while half of them had three confirmation gazes, which shows that confirmation gazes were very common. 

These differences in behaviour can be utilized in multimodal referencing in several ways. For example, pointing can be used as the main modality and could be tracked at all times, while gaze could be tracked only for a short time window before and after pointing. This is because almost all participants did not gaze during pointing at all or only gazed once. The mean modality delta time and confirmation time (1.91 and 2.43 seconds, respectively) can be used to determine this time frame for gaze tracking.

\subsubsection{Intraclass correlations}
Since users point at objects in a unique way generally~\cite{brown2016exploring,brown2014performance} and while driving~\cite{rumelin2013free}, intraclass correlations (ICCs) were calculated to assess the variance in the data originating from participants' differences. 
A fifth of the data variance originates from participants.
Some variables such as \textit{pointing duration}, \textit{pointing time} or \textit{modality delta time} had especially high ICCs. This indicates that individual differences influenced how long participants pointed at a building, how long it took to find the building and how long it took to start pointing.
Other variables (e.g. \textit{detection time}, \textit{confirmation time} and \textit{gaze confirmation duration}) had a relatively low ICC, indicating a low influence of inter-individual differences and homogeneous variations across participants. This could also be utilized in the referencing and tracking process. For example, the threshold for pointing detection could be lowered for users with shorter pointing duration, which would lead to better pointing accuracy.

\subsubsection{Statistical inference}
Inferential statistical analysis shows that the first hypothesis was supported. Participants performed better (i.e. pointed and gazed more accurately and pointed faster) in non-dense conditions compared to dense conditions with a large effect size.
There was no significant difference in distance, as well as no significant interaction between density and distance in terms of performance. Thus, hypothesis two was not supported, which could be attributed to the fact that most participants already pointed when the PoI was still far away, as previously mentioned.
Hypothesis three was partially supported since only gaze accuracy was significantly better during autonomous driving. 

Moreover, there was a significant difference for PoI orientation. 
PoIs on the right side of the road were gazed at more accurately and pointed at faster compared to buildings on the left side, which supports hypothesis four. 
However, this could be due to a learning effect (since more buildings were on the right side). Other possible reasons could be that most of the participants were right-handed or mainly drove on the right side of the road. Thus, they might have an attentional focus bias towards the right side. Furthermore, participants were instructed to point with their right hand even when pointing left, to make sure that their gesture was not out of range for the camera. This instruction might have negatively influenced pointing performance for left-side PoIs.

Hypothesis five was supported by both the descriptive statistics and the statistical inference where gaze accuracy was significantly better than pointing accuracy. However, this finding needs to be interpreted with caution, since gaze accuracy calculation in this approach still depended on the pointing modality and therefore cannot be considered as an independent modality. The results for these hypotheses would determine the way each modality is tracked and fused. For example, the results for hypothesis four could be utilized by the system through tracking both pointing and gaze modalities for a right-oriented PoI, while tracking only the pointing modality for a left-oriented PoI.

\subsubsection{Clustering}

Another approach for utilizing behavioural differences in fusion is clustering the participants. Clustering can be done using performance-related dependent variables and time-related dependent variables as features.
Clustering using performance-related dependent variables shows that there are too many clusters and it is hard to cluster participants based on pointing and gaze accuracy alone. However, a possible fusion approach would be modality switching (see~\autoref{fig:distraccuracywithquadrants}), where the system tracks pointing modality alone for users with high pointing accuracy and low gaze accuracy while it tracks gaze modality alone for users with low pointing accuracy and high gaze accuracy. As for time-related dependent variables, the clustering output looks more separable than that of the performance-related one; however, it is harder to interpret and apply to actual use cases since it depended on multiple dependent variables. An alternative approach is to heuristically cluster participants based on each dependent variable separately to enhance the pointing and gaze gestures tracking. For example, users that usually point for a short amount of time (i.e. \textit{pointing time} is small) should have a lesser time threshold for accurate pointing detection than users who point for a longer time, and so on.

\subsection{Limitations}

Finally, there are several limitations for the current work that we will address in future studies. First, the traditional feature extraction approach used in our approach involved several assumptions and transformation steps that propagated small rounding errors. This led to an error margin of $\pm 4$ degrees in pointing modality and $\pm 3$ degrees in gaze modality. Secondly, although the driving route was long to assure safety-critical and dynamic criteria, it was simple and easy by design to increase internal validity, which could lead to a lower external validity. Lastly, from a technical side, the eye gaze tracker lacked a real-time communication feature and did not allow for online synchronisation with other devices. This led to the use of an offline synchronisation approach which resulted in a timing error margin of $\pm 300$ milliseconds. Besides, both the eye gaze and pointing gesture trackers had several hardware and software limitations that led to 40\% of the recorded data being unusable.

\section{Conclusion and Future work}

In conclusion, person-specific behaviour can be exploited to enhance the referencing task performance using several different approaches such as:
triggering gaze tracking based on the pointing starting time; adjusting pointing tracking threshold based on user's pointing duration; changing tracking methodology based on target orientation; switching between the pointing and gaze tracking system based on users' clusters to maximize the overall referencing performance.

To overcome this work's limitations in future studies, we plan to use a deep learning approach for feature extraction, which would significantly reduce the traditional method's error margin. However, it would significantly reduce the explainability of the extracted feature as well, and the right hyperparameters for such an approach could be hard to find. 
Moreover, we plan to use more complicated and harder driving routes in further studies to increase the external validity. However, a more challenging driving task could slightly alter the user's behaviour. Finally, more reliable hardware that also supports online synchronisation could be used to avoid offline synchronization problems and output more reliable data. However, such devices can be significantly more expensive.

\begin{acks}
This work is partially funded by the German Ministry of Education and Research (project TRACTAT: Transfer of Control between Autonomous Agents; grant number 01IW17004).
\end{acks}

\bibliographystyle{ACM-Reference-Format}
\balance
\bibliography{sample-base}










\end{document}